\documentclass[10pt,letterpaper]{article}
\usepackage{opex3}
\usepackage{graphicx}
\usepackage{float}
\usepackage{amsmath,amssymb,latexsym}
\usepackage{bm}

\begin{document}

\title{Spatial interference of light: transverse coherence and the Alford and Gold effect}

\author{Jefferson Fl\'orez,$^{1,*}$ Juan-Rafael \'Alvarez,$^1$ Omar Calder\'on-Losada,$^1$ Luis-Jos\'e Salazar-Serrano,$^{1,2}$ and Alejandra Valencia$^1$}

\address{$^1$Laboratorio de \'Optica Cu\'antica, Universidad de los Andes, A.A. 4976, Bogot\'a, D.C., Colombia\\
$^2$ICFO-Institut de Ciencies Fotoniques, Mediterranean Technology Park, 08860 Castelldefels, Barcelona, Spain}

\email{$^*$j.florez34@uniandes.edu.co} 



\begin{abstract}
We study the interference between two parallel-propagating Gaussian beams, originated from the same source, as their transverse separation is tuned. The interference pattern as a function of such separation lead us to determine the spatial coherence length of the original beam, in a similar way that a Michelson-Morley interferometer can be employed to measure the temporal coherence of a transform limited pulse. Moreover, performing a Fourier transform of the two-beam transverse plane, we observe an intensity modulation in the transverse momentum variable. This observation resembles the Alford and Gold Effect reported in time and frequency variables so far. 
\end{abstract}




\section{Introduction}
\begin{figure}[htbp]
\begin{center}
\includegraphics[width=\textwidth]{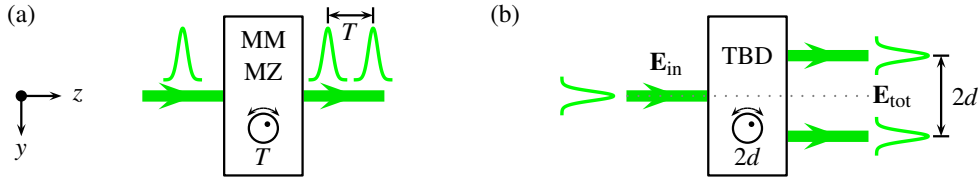}
\caption{Working principle of (a) a Michelson-Morley (MM) or a Mach-Zender (MZ) interferometer, and (b) a tunable beam displacer (TBD). In (a), an incoming pulse is divided longitudinally into two pulses, one of them delayed a tunable interval $T$ with respect to the other. In (b), the incoming beam is split into two beams that propagate in parallel, each one shifted by a transverse distance $d$ with respect to the incoming beam. The total tunable separation between the beams is $2d$.}
\label{TBD-Sketch}
\end{center}
\end{figure}

Interference of coherent light occurs if there is a non-random phase difference between two or more beams. For example, in a Michelson-Morley (MM) or in a Mach-Zender (MZ) interferometer, the two-beam relative phase is introduced by means of a path difference between the interferometer arms. The variation of this path difference allows to tune the temporal delay between the beams, as shown in Fig. \ref{TBD-Sketch}(a), resulting in an interference pattern when the interferometer output intensity is observed as a function of the temporal delay. For these interferometers, the maxima and minima on the intensity appear if the temporal delay is less or approximately equal to the temporal coherence of the light source, which in turn can be used to quantify its coherence time if we are dealing with a transform-limited pulse \cite{Goodman}. Furthermore, it is interesting to notice that interference can be observed as well by performing a spectral analysis of the output electric field \cite{Zou92}. In this second case, a periodic modulation on the output intensity is obtained as a function of the frequency variable for a fixed temporal delay that is greater than the coherence time of the source. The latter fact was named as the Alford and Gold Effect (AGE) by L. Mandel \cite{Mandel62}, originally measured by W. P. Alford and A. Gold \cite{AG58}, and expressed in terms of light interference by M. P. Givens \cite{Givens61}. The AGE has been observed for single photons \cite{Zou92} and for a pulsed laser \cite{Salazar14} using a MM interferometer.

Apart from the temporal domain, interference can also be observed in the spatial variables. For example, in the Young's double-slit experiment, the separation between the two slits must be less or approximately equal to the spatial coherence length of the light source in order to observe a bright and dark fringe pattern \cite{Goodman}. Indeed, the fringe visibility of a double-slit interferometer gives a measure of the degree of spatial coherence for a finite-size light source \cite{Redding11,Barcik13}. Although the Young's double-slit experiment is the usual choice to determine the spatial coherence of light, one may think on an alternative approach analogous to the way in which temporal coherence is quantified via a MM interferometer. In this paper, we achieve this goal by using a tunable beam displacer (TBD) similar to the one reported in \cite{Salazar15}. A TBD is a device that splits a light beam into two parallel-propagating beams, each one shifted by a tunable transverse distance $d$ with respect to the incoming beam, as shown in Fig. \ref{TBD-Sketch}(b). The capability of tuning the separation between the two beams allows us to observe a spatial interference pattern as a function of $d$, and therefore to have an analogous to the MM interference pattern as a function of the temporal delay. Additionally, by measuring the TBD output intensity at the Fourier plane, we observe a periodic intensity modulation similar to the temporal AGE with light pulses \cite{Givens61,Salazar14}. The experimental results are in agreement with a direct correspondence between the temporal and spatial interference patterns measured by means of a MM or a MZ interferometer and a TBD device, respectively.

\section{Theory}

%
%

To put on solid ground the analogy between the TBD and the MM interferometer, let us consider the situation depicted in Fig. \ref{TBD-Sketch}(b), in which a monochromatic, diagonally polarized, well-collimated Gaussian beam enters to the TBD. The transverse momentum distribution $\mathcal{E}_\text{in}(q_x,q_y)$ for this incident beam is given by a Gaussian function centered at $\mathbf{q}_0=(q_{0x},q_{0y})$,
\begin{equation}
\mathcal{E}_{\rm in}(q_x,q_y)\propto \exp\left\{-\frac{w_0^2}{4}\left[(q_x-q_{0x})^2+(q_y-q_{0y})^2\right]\right\},
\label{Fin}
\end{equation}
where $w_0$ is the beam waist. Then, the transverse component $\mathbf{E}_{\rm in}(x,y)$ of the electric field in position variables of such a beam, corresponds to the Fourier transform of $\mathcal{E}_{\rm in}(q_x,q_y)$, and is equal to
\begin{equation}
\mathbf{E}_{\rm in}(x,y)=E_0\exp\left(-\frac{x^2+y^2}{w_0^2}\right)\exp(iq_{0y}y)\hat{\mathbf{e}}_D,
\label{Ein}
\end{equation}
with $E_0$ the electric field amplitude, $\hat{\mathbf{e}}_D$ the polarization vector chosen to be diagonal, and $(x,y)$ the conjugate vector of the transverse momentum $(q_x,q_y)$. In the last expression, we have taken $q_{0x}=0$ since we are only interested in fields shifted along the $y$-direction. Note that the factor $\exp(iq_0y)$ comes from considering a small beam deviation in the $y$-direction with respect to the $z$-axis, which is defined perpendicular to the TBD input plane \cite{Saleh00}.

At a fixed transverse plane after the TBD, the total electric field $\mathbf{E}_{\rm tot}(x,y;d)$ is proportional to the superposition of two fields shifted by distances $+d$ and $-d$, respectively. This is
\begin{equation}
\mathbf{E}_{\rm tot}(x,y;d)\propto \mathbf{E}_{\rm in}(x,y-d)+e^{i\phi}\mathbf{E}_{\rm in}(x,y+d),
\label{Eout}
\end{equation}
where $\phi$ is a relative phase between the two fields that appears due to the working principle of the TBD. The total intensity $\tilde{I}_{\rm tot}(x,y;d)$ is then given by $\tilde{I}_{\rm tot}(x,y;d)= \parallel \mathbf{E}_{\rm tot}(x,y;d)\parallel^2.$

\begin{figure}[h]	
\centering
\includegraphics[scale=0.6]{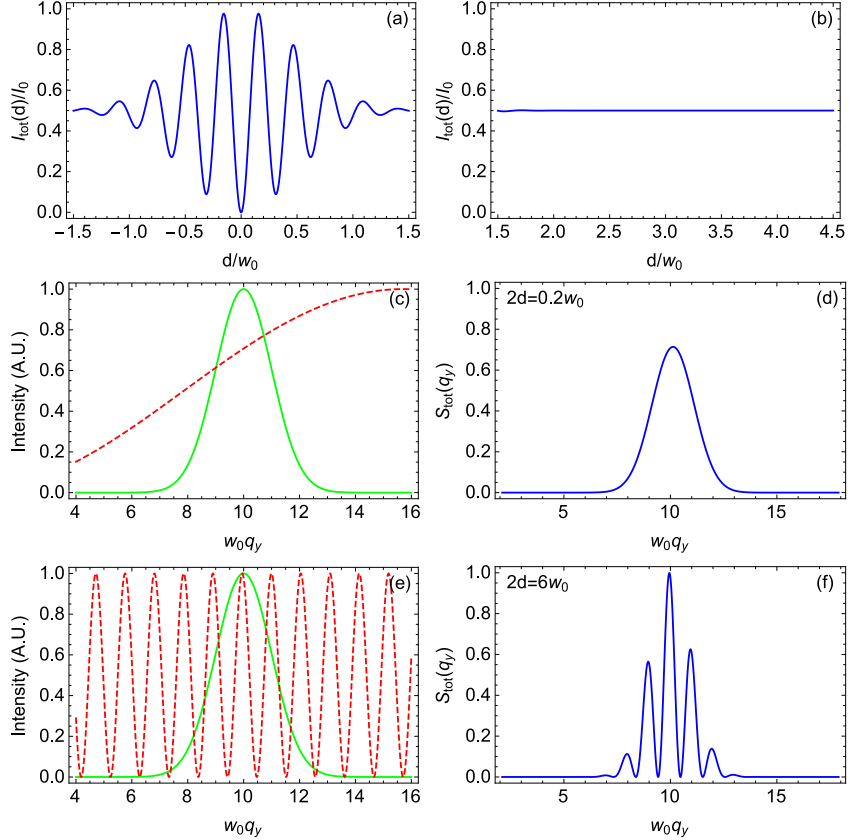}
\caption{Theoretical prediction for the interference in position, (a) and (b), and momentum variables, (d) and (f), of two parallel propagating beams after a tunable beam displacer. For (a) and (b), $q_0w_0=10$. In (c) and (e),  $S_\text{in}(q_y)$ is represented as a solid line; as dashed lines, the term $[1+\cos(2q_yd+\pi)]/2$ is plotted for two different periods, $\pi/d=1/(0.1w_0)$ and $1/(3w_0)$, respectively.}
\label{Theory}
\end{figure}

When one is interested in the temporal coherence and uses a MM interferometer to measure it, one must integrate the total intensity with respect to time since the detectors are not fast enough to resolve the temporal changes on such intensity \cite{Diels}. Spatially, we proceed in an analogous fashion by integrating the total intensity $\tilde{I}_{\rm tot}(x,y;d)$ over the detection area $A$.
Following this, the integrated intensity $I_{\rm tot}(d)$ reads
\begin{equation}
I_{\rm tot}(d)\propto\int_A dxdy \tilde{I}_\text{tot}(x,y)=\frac{|E_0|^2}{2}\left[1+\exp\left(-\frac{2d^2}{w_0^2}\right)\cos(2q_{0y}d+\phi)\right].
\label{Itot}
\end{equation}
From Eq. (\ref{Itot}), one recognizes the usual interference pattern of partially coherent beams mediated by the transverse displacement $d$. In analogy with the temporal interference in a MM interferometer, the Gaussian factor accompanying the cosine function is proportional to the degree of spatial coherence of the source \cite{Scully97}. It is then possible to recognize $\Delta y_\text{FWHM}=w_0\sqrt{2\ln 2}$ as the corresponding spatial coherence length. In particular, if $d$ is smaller or approximately equal to $w_0$, the interference is revealed as an intensity modulation governed by the product of the cosine function and an envelope given by the Gaussian term. In contrast, if $d$ is greater than $w_0$, one gets a constant total intensity equal to $I_0/2$ since the exponential function vanishes. We represent these behaviors in Figs. \ref{Theory}(a) and \ref{Theory}(b), respectively. A negative distance $d$ in Fig. \ref{Theory}(a) means that the lower beam after the TBD in Fig. \ref{TBD-Sketch}(b) has moved to the upper position, and the contrary for the other outcome of the TBD. The minimum in the same Fig. appears for $d=0$ due to internal reflections inside the TBD, which leads to $\phi=\pi$.

Apart from observing the spatial interference, the TBD also allows to measure the spatial AGE. To observe this formally, one needs to calculate the intensity $S_{\rm tot}(q_x,q_y)= \mathcal{E}_{\rm tot}^2(q_x,q_y)$ at the Fourier plane, where $\mathcal{E}_{\rm tot}(q_x,q_y)$ is the Fourier transform of the total electric field magnitude,
\begin{equation}
\mathcal{E}_{\rm tot}(q_x,q_y)\propto \int dxdy\ \parallel\mathbf{E}_{\rm tot}(x,y;d)\parallel\exp(-iq_xx-iq_yy).
\label{Fout}
\end{equation}
Since the effects of the TBD will only be observed in the $y$-direction, we take $q_x=0$, so that $S_{\rm tot}$ becomes
\begin{equation}
S_{\rm tot}(q_y)=\frac{S_\text{in}(q_y)}{2}\left[1+\cos(2q_yd+\phi)\right],
\label{Stot}
\end{equation}
with $S_\text{in}(q_y)=\mathcal{E}_{\rm in}^2(0,q_y)$, and $\mathcal{E}_{\rm in}$ given by Eq. (\ref{Fin}).

Observing Eq. (\ref{Stot}), one notices again an intensity modulation of $S_\text{in}(q_y)$ mediated by the cosine function, but in the conjugate variable $q_y$. In particular, if the period of oscillation $\pi/d$ of the cosine function is greater than the width $1/w_0$ of $S_\text{in}(q_y)$ (Fig. \ref{Theory}(c)), there is no modulation, as depicted in Fig. \ref{Theory}(d). In contrast, if $\pi/d<1/w_0$ (Fig. \ref{Theory}(e)), there is a modulation of the incoming beam momentum distribution, as shown in Fig. \ref{Theory}(f).

Concretely, from Figs. \ref{Theory}(d) and \ref{Theory}(f), we can see that as $d$ grows and becomes greater than $w_0$, a set of oscillations appear within the Gaussian envelope $S_\text{in}(q_y)$. This means that even if the two gaussian beams are far enough from each other in such a way that there is no interference pattern in the position variable, they do display an intensity modulation at the Fourier plane, revealing a spatial analogous to the AGE.

\section{Experimental implementation and Results}\label{Experimental}
\begin{figure}[htbp]
\begin{center}
\includegraphics{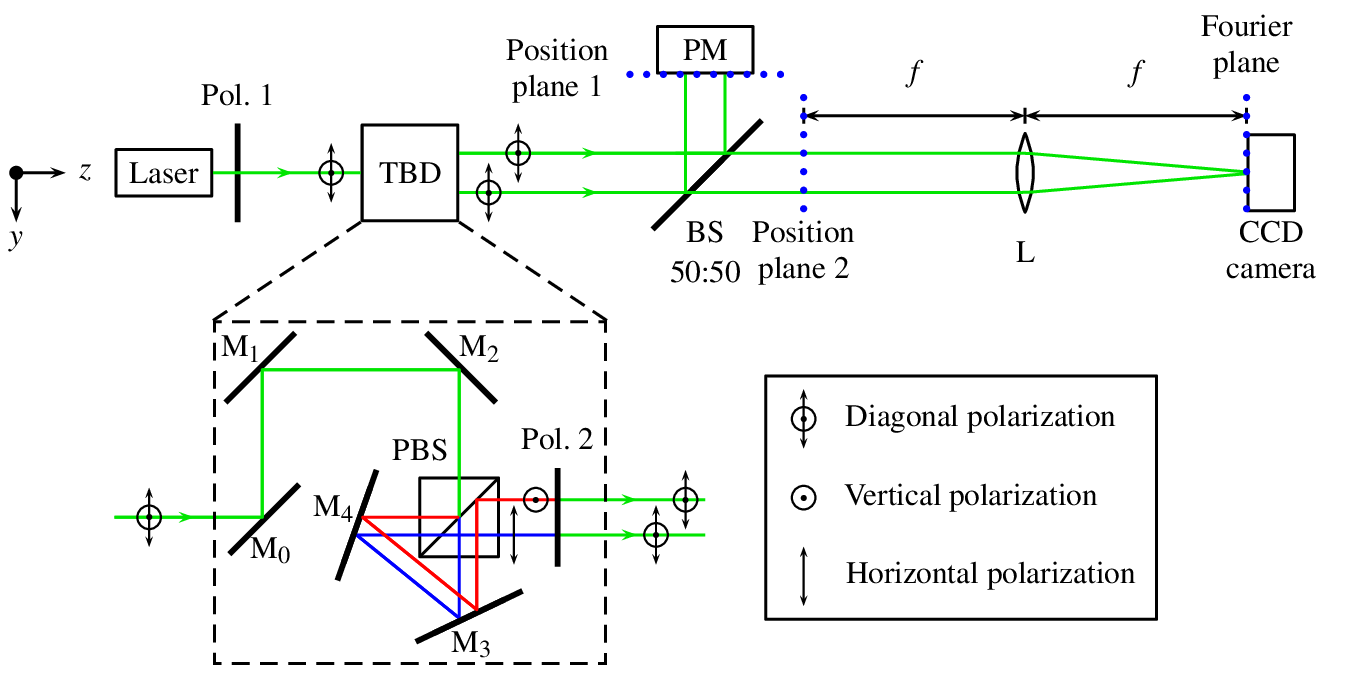}
\caption{Experimental setup to register the spatial interference of two Gaussian beams at the position and Fourier planes. M$_i$, with $i=0,1,...,4$, denote mirrors, Pol. 1 and Pol. 2 polarizers, TBD a tunable beam displacer, PM a power meter, PBS (BS) a polarizing (non-polarizing) beam splitter, and L a $f=75$ cm-biconvex lens. Position plane 2 has exactly the same information as Position plane 1. A detailed sketch of the TBD is shown in the inset.}
\label{TBD}
\end{center}
\end{figure}

To test the previous theory, we implemented the experimental setup shown in Fig. \ref{TBD}. Using a 808 nm-CW laser (Thorlabs, CPS808), we prepared $\mathbf{E}_\text{in}(x,y)$ with $w_0=(0.87\pm0.01)$ mm by means of a single mode fiber, and set the polarization of the beam to be diagonal by means of a polarizer (Pol. 1.) The TBD consisted of a polarizing beamsplitter (PBS), two mirrors (M$_3$ and M$_4$) mounted on a rotational stage, and a polarizer at $45^\circ$ (Pol. 2) to have the two output beams with the same polarization and intensity, as shown in the inset of Fig. \ref{TBD}. The mirrors M$_0$, M$_1$, and M$_2$ are placed in order to have both the TBD input and output along the $z$-direction.

\begin{figure}[h]	
\centering
\includegraphics[scale=0.6]{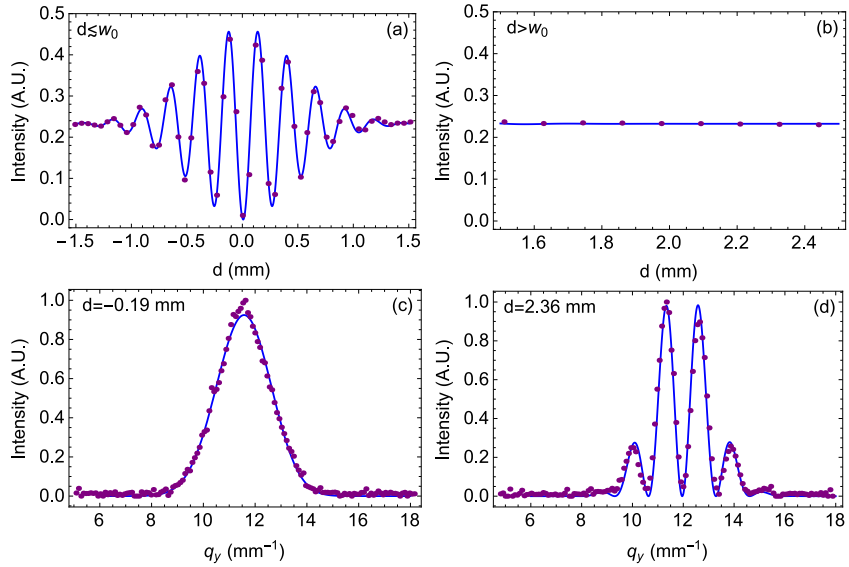}
\caption{Experimental results (circles) for the interference of two parallel-propagating beams in position and momentum variables. Interference in position as a function of the transverse displacement $d$ of the beams for (a) $d\lesssim w_0$, and (b) $d> w_0$. Interference at the Fourier plane as a function of the transverse momentum $q_y$ when (c) the two beams almost overlap ($d\approx 0$), and (d) $d=2.36$ mm. Error bars are completely within the experimental point size.}
\label{Results}
\end{figure}

In order to observe the total intensity in the position variable and in its conjugate at the same time, we placed a 50:50 beam splitter and measured in its reflection the intensity in position $I_\text{tot}(d)$, and in its transmission the intensity in momentum, $S_\text{tot}(q_y)$. The measurement in the position variable was done using a power meter (Coherent, BeamMaster-USB BM-7 Si-Enhanced), at what we call the Position plane 1. 
The results of this measurement are shown in Fig. \ref{Results}(a) and \ref{Results}(b), for $d\lesssim w_0$ and $d>w_0$, respectively. In Fig. \ref{Results}(a), a typical interference pattern is clearly observed as a function of the separation $d$. By fitting the data to Eq. (\ref{Itot}), the full width at half maximum (FWHM) of the gaussian envelope is $(1.12\pm0.03)$ mm, indicating the spatial coherence length of our light source. This result is in agreement with the expected theoretical value $\Delta y_\text{FWHM}=w_0\sqrt{2\ln 2}=(1.03\pm 0.01)$. In sharp contrast, Fig. \ref{Results}(b) shows no interference, represented as a flatline. The continuous lines in Fig. \ref{Results}(a) and \ref{Results}(b) correspond to the ones depicted in Fig. \ref{Theory}(a) and \ref{Theory}(b) for the theoretical model, respectively, and demonstrate that it is possible to observe an interference pattern spatially analogous to the one obtained in a MM interferometer.

The measurement of $S_\text{tot}(q_y)$ was done using a 2$f$-system composed by a biconvex lens of focal length $f=75$ cm, and placing a CCD camera (SBIG, ST-1603ME) at the Fourier plane. The 2$f$-system transforms the position variables, in what we call Position plane 2, into the corresponding conjugate ones. In the Fourier plane there is a one-to-one correspondence between $q_y$ and the $y$ coordinate $y_\text{FP}$ at the Fourier plane, given by $q_y=2\pi y_\text{FP}/(\lambda f)$, with $\lambda$ denoting the wavelength of the light beam. The experimental results are shown in Fig. \ref{Results}(c) and \ref{Results}(d) for $d<w_0$ and $d>w_0$, respectively. From Fig. \ref{Results}(c), it is observed that at the Fourier plane there is no intensity modulation for $d$ close to zero, retrieving $S_\text{in}(q_y)$, as expected. However, in Fig. \ref{Results}(d) there is a modulation of $S_\text{in}(q_y)$. These observations are analogous to the AGE reported in temporal variables so far, and exhibit a similarity between the temporal and spatial degrees of freedom of light. It is interesting to notice that the center of both Figs. \ref{Results}(c) and \ref{Results}(d) is $(11.96\pm0.03)$ mm$^{-1}$, indicating that this is the value of $q_{0y}$ in Eq. (\ref{Ein}). Physically, this value is related to a small deviation of the incoming beam with respect to $z$-axis, given by an angle $\theta=q_{0y}\lambda/2\pi$. Using our measured $q_{0y}$, we get $\theta=(1.528\pm0.004)$ mrad.

\begin{figure}[h]	
\centering
\includegraphics[scale=0.6]{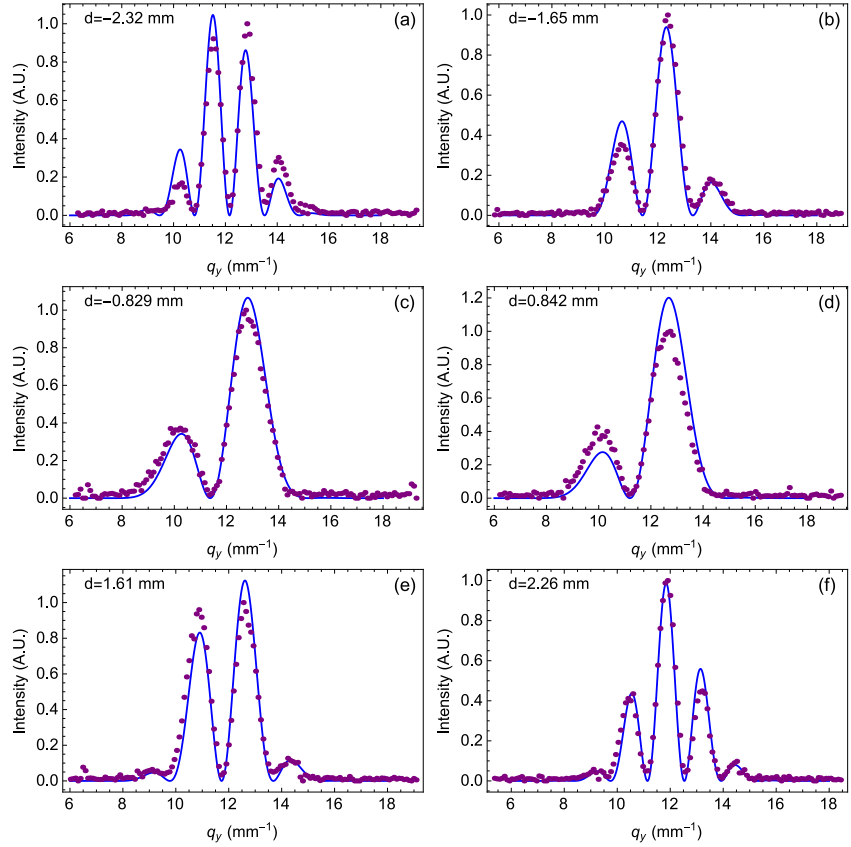}
\caption{Experimental results (circles) for the interference at the Fourier plane as the separation distance  between the beams varies. The solid (blue) lines are the theoretical predictions according to Eq. (\ref{Stot}).}
\label{Results-q}
\end{figure}

As complementary measurements, we report in Fig. \ref{Results-q}, the intensity at the Fourier plane $S_\text{tot}(q_y)$ for various separations $d$ between the beams. From these graphs, one observes how the modulation of $S_{\text{in}}(q_y)$ varies according to the separation $d$ between the beams and sees that when $|d|$ increases, the number of peaks increases as well.

\section{Conclusions}
We have performed and observed interference of light in its spatial degree of freedom. This was done using a tunable beam displacer that allowed us to recreate in the spatial domain an interferometer analogous to the Michelson-Morley. In this way, we were able to measure the spatial coherence length of a light beam, and to observe the spatial Alford and Gold effect. Using the fact that there is a correspondence between the temporal and spatial degrees of freedom of light, our results open the door to new interferential experiments recreating the results obtained in the temporal regime.

\section*{Acknowledgments} 
We acknowledge useful discussions with David Guzm\'an. This work was supported by Facultad de Ciencias, Universidad de los Andes, Bogot\'a, Colombia, under project number P15.160322.009/01-02-FISI06.
\end{document}